\documentstyle[twoside,fleqn,espcrc2]{article}

\newcommand{\AmS}{{\protect\the\textfont2
  A\kern-.1667em\lower.5ex\hbox{M}\kern-.125emS}}

\def\inbar{\vrule height1.5ex width.4pt depth0pt}
\def\IB{\relax{\rm I\kern-.18em B}}
\def\IC{\relax\,\hbox{$\inbar\kern-.3em{\rm C}$}}
\def\ID{\relax{\rm I\kern-.18em D}}
\def\IE{\relax{\rm I\kern-.18em E}}
\def\IF{\relax{\rm I\kern-.18em F}}
\def\IG{\relax\,\hbox{$\inbar\kern-.3em{\rm G}$}}
\def\IH{\relax{\rm I\kern-.18em H}}
\def\II{\relax{\rm I\kern-.18em I}}
\def\IK{\relax{\rm I\kern-.18em K}}
\def\IL{\relax{\rm I\kern-.18em L}}
\def\IM{\relax{\rm I\kern-.18em M}}
\def\IN{\relax{\rm I\kern-.18em N}}
\def\IO{\relax\,\hbox{$\inbar\kern-.3em{\rm O}$}}
\def\IP{\relax{\rm I\kern-.18em P}}
\def\IQ{\relax\,\hbox{$\inbar\kern-.3em{\rm Q}$}}
\def\IR{\relax{\rm I\kern-.18em R}}
\def\ZZ{\relax{\sf Z\kern-.4em Z}}
\def\fnote#1#2{\begingroup\def\thefootnote{#1}\footnote{#2}\addtocounter
{footnote}{-1}\endgroup}
\def\beq{\begin{equation}}
\def\eeq{\end{equation}}
\def\bea{\begin{eqnarray}}
\def\eea{\end{eqnarray}}
\def\lleq#1{\label{#1}\eeq}
 \def\llea#1{\label{#1}\eea}
\let\nn=\nonumber

\def\notin{\ \hbox{{$\in$}\kern-.51em\hbox{/}}}

\def\a{\alpha}

   \def\th{\theta}
  \def\cC{{\cal C}}

\def\lra{\longrightarrow}
\def\lolra{\longleftrightarrow}

\def\hbar{\bar h}

 \def\II{{\bf II}}

\title{Heterotic/Type II Duality in D$=$4 and String-String
       Duality} 

\author{
        B.Hunt\address{SFB 343, Postfach 100131, 33501 Bielefeld, FRG}%
        \thanks{sfb2@mathematik.uni-bielefeld.de}
        , 
        M.Lynker\address{Department of Physics and Astronomy, 
                         Indiana University South Bend, \\  
        1700 Mishawaka Ave., South Bend, IN 46634, USA}%
        \thanks{mlynker@siggy.iusb.edu}  
        and 
        R.Schimmrigk\address{Physikalisches Institut, 
                             Universit\"at Bonn, 
                             53115 Bonn, FRG}%
         \thanks{netah@avzw02.physik.uni-bonn.de}}

\begin{document}
\begin{titlepage}
\begin{large}
\begin{flushright} 
hep-th/9609082

BONN-TH-96-08 
\end{flushright}

\vskip .5truein
\centerline{\Large\bf Heterotic/Type II Duality in D$=$4 and String-String
       Duality\fnote{\diamond}{Based on a talk given
                           at the Fourth International
              Conference on Supersymmetry (SUSY '96), 
           College Park, May 1996}}

\vskip .6truein
\centerline{\sc B.Hunt$^1$\fnote{\star}{sfb2@mathematik.uni-bielefeld.de},
      M.Lynker$^2$\fnote{\ddagger}{mlynker@siggy.iusb.edu}
   and R.Schimmrigk$^3$\fnote{\dagger}
                        {netah@avzw02.physik.uni-bonn.de}}

\vskip .3truein
\centerline{\it $^1$SFB 343, Postfach 100131, 33501 Bielefeld, FRG}
\vskip .2truein
\centerline{\it $^2$Department of Physics and Astronomy,
                  Indiana University South Bend}
\centerline{\it 1700 Mishawaka Ave., South Bend, IN 46634, USA}%
\vskip .2truein
\centerline{\it $^3$Physikalisches Institut, Universit\"at Bonn}
\centerline{\it Nussallee 12, 53115 Bonn, FRG}

\vskip .8truein
\centerline{\bf ABSTRACT}
\vskip .1truein 
\noindent
We discuss the structure of Heterotic/Type II duality in four
dimensions as a consequence of string-string duality in six
dimensions. We emphasize the new features in four dimensions
which go beyond the six dimensional vacuum structure and
pertain to the way particular K3 fibers can be embedded in
Calabi-Yau threefolds. Our focus is on hypersurfaces as well
as complete intersections of codimension two which arise via
conifold transitions.
\end{large}

\end{titlepage}

\vfill \eject 

\begin{abstract}
We discuss the structure of Heterotic/Type II duality in four 
dimensions as a consequence of string-string duality in six 
dimensions. We emphasize the new features in four dimensions 
which go beyond the six dimensional vacuum structure and 
pertain to the way particular K3 fibers can be embedded in 
Calabi-Yau threefolds. Our focus is on hypersurfaces as well 
as complete intersections of codimension two which arise via 
conifold transitions.
\end{abstract}

\maketitle

\section{Introduction}

Over the last two years a large variety of duality conjectures 
has been put forward, relating different compactifications of 
the basic five string theories. It clearly would be helpful
if all these dualities 
\cite{ht95,ew95,kv95,klm95,vw95,het2,mth,fth} could be 
traced back to a few basic
ancestors. Progress in this direction has been made in several 
articles \cite{basics}. In the present 
paper we focus on the Heterotic/Type II duality in D=4 
\cite{kv95} as a consequence of string-string duality in D=6. 
The expectation that such a relation between dualities might hold 
is based on the recognition \cite{klm95} 
that this duality in D=4 is possible for K3-fibered Calabi-Yau 
threefolds. Thus one might expect to be able to apply 
string-string duality fiber-wise in the limit when the fiber 
varies little - the adiabatic limit of \cite{vw95}. The
concrete implementation of this idea \cite{hs95}, which we will 
describe in the following, shows that there
are a number of new twists that introduce additional structure 
beyond that of the K3 in D=6.
 
\section{String-String Duality in D=6}

It was observed in \cite{ns88} 
that the heterotic string compactified on 
the four-torus T$^4$ has the same moduli space 
SO(20,4;$\ZZ){\big \backslash}$SO(20,4;$\IR){\big /}$
SO(20;$\IR)\times $ SO(4;$\IR$) 
as the type IIA string compactified on the K3 surface. Thus 
the compactification of two different string theories on the 
only two Calabi-Yau complex surfaces turn out to be identical.
Dualities in 4D then follow by compactifying these vacua 
further on 2-tori \cite{ht95,as94}.
 
\section{Heterotic/Type II Duality in D=4}

Evidence for a more general dualities in four dimensions 
has been presented in \cite{kv95}. It was shown there that 
Calabi-Yau threefolds exist for which one can find bundles 
$V_i$ on K3$\times $T$^2$ such that after Higgsing one finds 
the relation
\beq
{\rm Het(K3}\times {\rm T}^2) \lolra {\rm IIA(CY}_3).
\eeq
With hindsight such a relation might have been expected for 
a particular class of Calabi-Yau threefolds. By multiplying 
T$^4 \lra $T$^4\times \IP_1$ and K3 $\lra $K3$\times \IP_1$ 
and viewing this as the {\it local} structure of the 
threefolds, which has to be twisted such as to make these 
into fibered Calabi-Yau manifolds, one might hope to establish 
this duality in four dimensions. 

A concrete procedure to do this twisting has been described 
in \cite{hs95}. This construction shows to what extent 
it is in fact the K3 structure which is responsible for the 
duality in 4D. It also explicates the new features that arise when 
string-string duality is lifted to from D=6 to D=4. 
We now review some of the salient properties  
of the relevant twist map.

\subsection{The Twist Map}

The starting point of the construction of \cite{hs95} is a  
K3 surface with an automorphism 
$\ZZ_{\ell} \ni {\bf m}_{\ell}$. Depending on the order of 
this cyclic symmetry 
group we choose a Riemann surface $\cC_{\ell}$ of genus 
$g(\ell)$ and the projection $\pi_{\ell}: \cC_{\ell}\lra \IP_1$. 
The twist map    
\beq
\cC_{\ell} \times K3 {\Big /}\ZZ_l \ni 
             \pi_{\ell}\times {\bf m}_{\ell} 
\lra {\rm CY}_3 
\eeq
then produces the explicit twisting. 

For the class of weighted Calabi-Yau hypersurfaces this map 
takes the following explicit form. Given a K3 
surface $\IP_{(k_0,k_1,k_2,k_3)}[k]$ with $k/k_0 = \ell \in \IN$ 
we define the curve $\cC_{\ell}=\IP_{(2,1,1)}[2\ell]$ of 
genus $g(\ell)=(\ell-1)^2$ and the map  
\bea
\IP_{(2,1,1)}[2\ell] \times \IP_{(k_0,k_1,k_2,k_3)}[k]
{\Big /} \ZZ_{\ell} ~~~~~~ &&\nn \\
 ~~~~~~~\lra ~~~ 
\IP_{(k_0,k_0,2k_1,2k_2,2k_3)}[2k]  &&  
\eea
via 
\bea 
((x_0,x_1,x_2),(y_0,\dots,y_3)) \lra ~~~~~~~~~&&    \nn \\  
 ~~~~~~~~
\left(x_1 \sqrt{\frac{y_0}{x_0}},x_2\sqrt{\frac{y_0}{x_0}},
                 y_1,y_2,y_3\right). &&   
\llea{twist}
This map then embeds the orbifold of the product on the lhs into 
the weighted four-spaces $\IP_{(k_0,k_0,2k_1,2k_2,2k_3)}$ as a 
hypersurface of degree $2k$.

\subsection{Properties of the Twist Map}

The map (\ref{twist}) shows that the structure of the 
   threefold is indeed determined by a single K3 surface, a 
   feature that supports the fiber-wise reduction of D=4 duality.
Using the results of \cite{ew95} it is in particular to be 
    expected that the degeneration structure of the K3 surface 
   will play an important role in the determination of the gauge 
   structure of the heterotic dual.    

The map also shows, however, the new features introduced by 
the twisting: the action of $\pi_{\ell}\times {\bf m}_{\ell}$ 
has fixed points which have to be resolved. This resolution 
introduces new cohomology and therefore the heterotic gauge 
structure is not completely determined by the K3 fiber. 

In the weighted category this aspect has two manifestations:
\begin{enumerate}
\item If $k_0=1$, as was the case in the original considerations 
    of \cite{klm95,het2,fth,mth}, the action of $\ZZ_{\ell}$ generates a 
    singular curve on the threefold which lives in the K3 fiber. 

      This curve, which is not present in the original K3 surface, 
      has the effect of introducing additional branchings in the 
      resolution diagram of the Calabi-Yau threefold. It is this 
      branching which determines the final gauge structure of the 
      heterotic dual of the IIA theory on the threefold.  
\item More generally one encounters $k_0>1$, considered for 
   threefolds in \cite{ls95} and for fourfolds in \cite{bs96}.
   In such a situation the orbifolding $\ZZ_{\ell}$ generates further 
    singularities on the threefold (and the fourfold). 
    Depending on the structure 
     of the weights these additional singularities can be either 
    points or additional singular curves (and surfaces for fourfolds).
\end{enumerate}

As an example which illustrates the first point 
we consider the K3 surface 
\beq
K=\{y_1^{42}+y_2^7+y_3^3+y_4^2=0\}\in
\IP_{(1,6,14,21)}[42].
\lleq{k3-42}
$K$ has an automorphism 
$\ZZ_{42}:~(y_1,y_2,y_3,y_4)\mapsto (\a y_1,y_2,y_3,y_4)$,
where $\alpha$ is a $42^{nd}$ root of unity.
Thus the associated curve $\cC_{42} = \IP_{(2,1,1)}[84]$ 
of the pair (K3, $\ZZ_{42}$) 
is a degree 42 cover of $\IP_1$, branched
at the 84 roots of $-1$ under the projection 
$\pi_{42}:\IP_{(2,1,1)}[84] \lra \IP_1$, 
the map $\pi_{42}$ is given explicitly by
$\pi_{42}(x_0,x_1,x_2)\mapsto (x_1,x_2)$. 
The action of $\ZZ_{42}$ on the product then has 84 fixed divisors, 
namely the copies of $K$ lying over the points 
$q_1,\ldots q_{84}\in \cC_{42}$ which is the
fixed point set of $\ZZ_{42}$ acting on $\cC_{42}$. These fixed
divisors
are the degenerate fibers of the K3 fibration of the quotient
$\ZZ_{42}\backslash \cC_{42}\times K$ over $\IP^1$ (given by
projection to
the first factor). Instead of resolving the singularities of the
quotient, we use the twist map (\ref{twist}). 
The image is the well-known Calabi-Yau $M\in
\IP_{(1,1,12,28,42)}[84]$. 
The degenerate fibers of $M$ as a K3 fibration are
cones over the curve 
$C=\{y_2^7+y_3^3+y_4^2=0\}$ in $\IP_{(6,14,21)}[42]$.
After resolution of the ambient projective space, 
the Euler number is $\chi(C)=11$. There is a
$\ZZ_2$, a $\ZZ_3$ and a $\ZZ_7$ fixed point on the curve, 
leading to 1, 2 and 6 new curves, respectively.
Thus we have $h^{1,1}(M)=2+9 =11$ and the Euler number is 
obtained from the fibration formula \cite{hs95} 
\beq
\chi(M)
= 2(1-\ell)\cdot 24 + 2\ell (\chi(C) + 2k_1 -1)
\lleq{fib-euler}
as $\chi=-960$.                     
Finally, to find the invariant part under the monodromy, 
we only have to determine the
Picard lattice of $K$. Our automorphism
group $\ZZ_{42}$ is the group denoted $H_K$ there, and this group
leaves precisely the Picard lattice invariant. Now we apply 
the results of Kondo \cite{sk92} which imply that 
$K$ is the {\it unique} K3 surface with a
$\ZZ_{42}$-automorphism, and that $K$ is an elliptic surface
(\ref{k3-42})
for which Kondo shows that the Picard lattice is $S_K=U\oplus E_8$.
Thus
this is the lattice of the gauge group of the heterotic dual. In
terms of
the curve $C$ this invariant lattice is described by the resolution
diagram
above, while in terms of the elliptic surface it is a union of a
section
and a singular fiber of type ${\bf II}^*$ (see \cite{hs95} for more 
details).

The twist map (\ref{twist}) furthermore makes explicit how a single  
K3 surface can lead to different threefolds: consider 
the K3 surface 
$\IP_{(1,2,3,3)}[9]\ni \{y_0^9+y_1^3y_2+y_2^3+y_3^3=0\}$
with the automorphism 
$\ZZ_9: (y_0,y_1,y_2,y_3) \longmapsto 
            (\a y_0,y_1,y_2,y_3)$.
Associated to this pair (K3, $\ZZ_9$) we choose the curve 
$\IP_{(2,1,1)}[18]$ of genus $g=17^2$. With these ingredients 
the twist map becomes 
\beq
\IP_{(2,1,1)}[18]\times \IP_{(1,2,3,3)}[9]/\ZZ_9 \lra
\IP_{(1,1,4,6,6)}[18] \nn
\eeq
leading to a weighted hypersurface with Hodge numbers
$(h^{(1,1)},h^{(2,1)})=(9,111)$.
Using the same K3 surface $\IP_{(1,2,3,3)}[9]$ of the previous 
example but now with a different automorphism 
$\ZZ_3: (y_0,y_1,y_2, y_3) \longmapsto
            (y_0,y_1,y_2, \a y_3)$ 
we are led instead to a different twist map
\beq
\IP_{(2,1,1)}[6]\times \IP_{(3,1,2,3)}[9]/\ZZ_3 \lra
\IP_{(3,3,2,4,6)}[18], \nn \\  
\lleq{lsex}
resulting in one of the K3 fibrations of \cite{ls95} with
spectrum $(h^{(1,1)},h^{(2,1)})=(5,53)$. Beyond the singular 
$\ZZ_2$-curve the twist map in this case also produces a 
second ($\ZZ_3-$) singular curve $\IP_{(1,1,2)}[6]$ on 
the threefold.
 
\section{Application to the Unification of Vacua}

\subsection{Weighted Conifold Transitions}  

In \cite{coniII} it was shown how conifold transitions, 
between Calabi-Yau manifolds, introduced in \cite{cdls88}, 
connect physically 
distinct vacua in type II string theory in a physically sensible way.
In \cite{ls95} such transitions were generalized to the framework 
of weighted Calabi-Yau manifolds \cite{weighted}. 
In general conifold transitions connect K3-fibered manifolds 
with spaces which are not fibrations. A simple example 
is the transition from the quasismooth octic $\IP_{(1,1,2,2,2)}[8]$   
to the quintic $\IP_4[5]$. It can be shown, however,  
that certain types of weighted conifold transitions 
exist which do connect K3 fibered manifolds \cite{ls95}.
A simple class of such conifold transitions  between 
fibered manifolds is provided by the weighted splits 
summarized in the diagram  
\bea 
\IP_{(2l,2l,2m,2k-1,2k-1)}[2(d+l)]~ \lolra ~~~~~~~~~~~~~&&  \nn \\ 
      \matrix{\IP_{(1,1)}\hfill \cr \IP_{(2l,2l,2m,2k-1,2k-1)}\cr}
       \left[\matrix{1&1\cr 2l&2d\cr}\right],&&    
\eea
where $d=(2k-1+l+m)$. 
Here the hypersurfaces, containing the K3 surfaces
$\IP_{(2k-1,l,l,m)}[2k-1+2l+m]$, split into codimension two manifolds
which contain the K3 manifolds
\beq
\matrix{\IP_{(1,1)}\hfill \cr \IP_{(l,l,m,2k-1)}\cr}
\left[\matrix{1&1\cr l&d\cr}\right]
\lleq{splitk3}
of codimension two.

\subsection{Twist Map for Split Manifolds}

In order to see the detailed fiber structure of the above 
varieties of codimension two it is useful to generalize the 
twist map of \cite{hs95} to complete intersection manifolds. 
Consider the K3 surfaces of the type (\ref{splitk3}) and the 
associated curves $\IP_{(2,1,1)}[2d]$ with $d=(l+m+2k-1)$. 
With these ingredients 
we can define a generalized twist map 
\bea 
\IP_{(2,1,1)}[2d] \times 
  \matrix{\IP_{(1,1)}\hfill \cr \IP_{(l,l,m,2k-1)}\cr}
  \left[\matrix{1&1\cr l&d\cr}\right]
 ~~~~~~~~~~~&&\nn \\
~~~~~~~\lra ~~~~
 \matrix{\IP_{(1,1)}\hfill \cr \IP_{(2l,2l,2k-1,2k-1,2m)}\cr}
       \left[\matrix{1&1\cr 2l&2d\cr}\right] && 
\eea
via 
\bea
\left((x_0,x_1,x_2),(u_0,u_1),(y_0,...,y_3)\right) 
                       \lra ~~~~~~~~~~&& \nn \\  
~~~~~\left((u_0,u_1),
         (x_1\sqrt{\frac{y_0}{x_0}},x_2\sqrt{\frac{y_0}{x_0}},
              y_1,y_2,y_3)\right) &&.
\eea
Again we see that the quotienting introduces additional singular 
sets and the remarks of Section 3.2 apply in the present context 
as well. In particular we see the new singular curve 
\beq 
\ZZ_2:~~C=\matrix{\IP_{(1,1)}\hfill \cr \IP_{(l,l,m)}\cr}
  \left[\matrix{1&1\cr l&d\cr}\right]
\eeq
which emerges on the threefold image of the twist map.

An example for such a transition between manifolds which are
K3-fibered as well as elliptically fibered is given by
\beq
\IP_{(1,1,2,4,4)}[12] ~\lolra ~
\matrix{\IP_{(1,1)}\hfill \cr \IP_{(4,4,1,1,2)}\cr}
\left[\matrix{1&1\cr 4&8\cr}\right] \nn \\   
\lleq{hetsplit}
where the Hodge numbers of the hypersurface are  
$(h^{(1,1)},h^{(2,1)})=(5,101)$ while those of the 
codimension two threefold $(h^{(1,1)},h^{(2,1)})=(6,70)$.  
Here the transverse codimension two variety of the rhs 
configuration is chosen to be 
\bea
p_1 &=& x_1y_1 + x_2y_2 \nn \\ 
p_2 &=& x_1(y_2^2+y_4^8+y_5^4) + x_2(y_1^2+y_3^8-y_5^4) 
\eea
which leads to the determinantal variety in the lhs 
hypersurface configuration 
\beq
p_{\rm det}=y_1^3 -y_2^3 +(y_1y_3^8 
      - y_2y_4^8)-(y_1+y_2)y_5^4.  
\lleq{detvar}
This variety is singular at 
$\IP_{(4,4,1,1,2)}[4~4~8~8] = 32$ nodes, which can be resolved 
by deforming the polynomial. 

Contained in these 2 CY--fibrations are the K3 configurations
\beq
\IP_{(2,2,1,1)}[6] ~\lolra ~
\matrix{\IP_{(1,1)}\hfill \cr \IP_{(2,2,1,1)}\cr}
\left[\matrix{1&1\cr 2&4\cr}\right],
\lleq{k3split2}
where the left-right arrow indicates that these two
are indeed related by splitting. The K3 of the rhs is
obtained by considering the divisor
\beq
D_{\th} = \{y_4=\th y_3\}
\lleq{div1}
which leads to
\beq
\matrix{\IP_{(1,1)}\hfill \cr \IP_{(4,4,1,2)}\cr}
\left[\matrix{1&1\cr 4&8\cr}\right],
\eeq
with
\bea
p_1&=&x_1y_1 + x_2y_2 \nn \\
p_2&=& x_1(y_2^2 + \th^8 y_3^8 + y_5^4)
         + x_2(y_1^2 + y_3^8 - y_5^4). \nn \\
\eea
Because of the weights in the weighted $\IP_3$ this is equivalent
to the rhs of (\ref{k3split2}) with
\bea
p_1 &=&x_1y_1 + x_2y_2 \nn \\
p_2 &=& x_1(y_2^2 + \th^8 y_3^4 + y_5^4)
         + x_2(y_1^2 + y_3^4 - y_5^4). \nn \\
\eea
The determinantal variety following from this space is
\beq
p_s = y_1^3-y_2^3 +(y_1 - \th^8 y_2)y_3^4 - (y_1-y_2)y_5^4.
\eeq
But this is precisely what one gets by considering the divisor
in the determinantal 3-fold variety (\ref{detvar}) and thus
we see that the conifold transitions take place in the fiber
of the CY 3-fold.

To elucidate the heterotic gauge structure of these K3 fibrations
it is necessary to determine the singularity structure of the fibers.
In the hypersurface on the lhs the singularities
\bea
\ZZ_2 &:& ~~\IP_{(2,2,1)}[6] \nn \\
\ZZ_4 &:& ~~\IP_1[4] = 3 {\rm pts}
\eea
are all contained in the K3 fiber. Similarly one finds for the
singularities of the codimension 2 space
\bea
\ZZ_2 &:& ~~ \matrix{\IP_{(1,1)}\hfill \cr \IP_{(2,2,1)}\cr}
\left[\matrix{1&1\cr 2&4\cr}\right]   \nn \\
\ZZ_4 &:& ~~ \matrix{\IP_1 \cr \IP_1\cr}
\left[\matrix{1&1\cr 1&2\cr}\right] = 3 {\rm pts}
\eea

Both singular curves thus lead to the resolution diagram
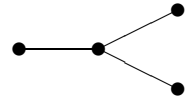
\begin{small}
\begin{center}
\begin{picture}(100,40)
\put(20,20){\circle*{5}}
\put(20,20){\line(1,0){30}}
\put(50,20){\circle*{5}}
\put(50,20){\line(2,1){30}}
\put(80,35){\circle*{5}}
\put(50,20){\line(2,-1){30}}
\put(80,5){\circle*{5}}
\end{picture}
\end{center}
\end{small}
which is the Dynkin diagram of SO(8).
                                                       
The heterotic dual of the type II theory on the hypersurface can 
be obtained by starting from the torus compactification 
\beq
(M^8\times T^2, E_8\times E_8\times SU(3)\times U(1)^2)
\eeq
with enhanced SU(3) symmetry point on the torus.
Compactifying further on a K3 with gauge bundles
\beq
V_i \lra K3,~~~i=1,2,3
\eeq
of rank $2,2,3$ respectively with
\beq
\int_{K3} c_2(V_i) \in \{10,8,6\}
\eeq
respectively and embedding the two SU(2) bundles in the two E$_8$s
respectively and the SU(3) into the SU(3) leads to the gauge group
$E_7\times E_7\times SU(3)\times U(1)^2$ for the 4--dimensional
theory and the numbers
\beq
N_{\bf 56}(V_1) = 3,~~~~
N_{\bf 56}(V_2) = 2
\eeq
for the two rank 2 bundles as well as the number of singlets
\beq
N_m(V_1) = 17,~N_m(V_2) = 13,~N_m(V_3) = 10, 
\eeq
which, together with the universal 20 of K3, leads to a total of
60 singlets.  The 4D theory thus is described by
$(M^4\times K3\times T^2, E_7 \times E_7\times U(1)^2)$ 
with $(n_H,n_V)=(60,16)$ and $3\cdot {\bf 56}$ in each of the E$_7$s.
Breaking the second E$_7$ down to SO(8) 
leads to  
$(M^4\times K3\times T^2, E_7 \times SO(8) \times U(1)^2)$ 
with 
$(n_H,n_V)=(67,13)$ and $3\cdot {\bf 56}$ in the surviving E$_7$.
Finally, breaking down the E$_7$ completely leads to
an additional $3\cdot 56-133 =35$  
singlets and therefore one ends up with a model
$(M^4\times K3\times T^2, SO(8) \times U(1)^2)$ 
with $(n_H,n_V)=(102,6)$ and no residual matter.
Taking into account the graviphoton and the axion-dilaton 
multiplet shows that this agrees with the Calabi-Yau 3fold 
cohomology.

\end{document}